# Simulation of Collision Resistant Secure Sum Protocol

Samiksha Shukla
Department of CSE
Christ University
Bangalore, India
samiksha.shukla@gmail.com

Dr. G. Sadashivappa
Department of Telecommunication
R V College of Engineering
Bangalore, India
sadashivappag@rvce.edu.in

Dr. Durgesh Kumar Mishra
Department of CSE
Sri Aurobindo Institute of Tech
Indore, India
drdurgeshmishra@gmail.com

*Abstract— secure multi-party computation is widely studied area in computer science. It is touching all most every aspect of human life. This paper demonstrates theoretical and experimental results of one of the secure multi-party computation protocols proposed by Shukla et al. implemented using visual C++. Data outflow probability is computed by changing parameters. At the end, time and space complexity is calculated using theoretical and experimental results.*

## I. INTRODUCTION

SMC works are mostly allocated to software agents, which can signify the participating parties, anonymizers and trusted third party. For example, Meeting scheduling problem where multiple employees will be the part of meeting but no one is interested in sharing his/her personal plan. Here, SMC techniques could be applied in such a way that meeting can be scheduled without any conflict and by hiding personal preferences.

The privacy preservation is a big challenge for data generated from various sources such as social networking sites, online transactions, weather forecast to name a few. Due to the socialization of the internet and cloud computing pica bytes of unstructured data is generated online with intrinsic values. The inflow of big data and the requirement to move this information throughout an organization has become a new target for hackers. This data is subject to privacy laws and should be protected. The proposed protocol is one step toward the security in case of above circumstances where data is coming from multiple participants and all the participants are concerned about individual privacy and confidentiality.

One technique to minimize privacy loss is to encode and hide [2] (identity) the association of the parties' sensitive data. Usually this is concerned as encryption and anonymization.

## II. RELATED WORKS

Various researchers are working in the field of computation security, few people proposed solution for secure sum and other SMC operations. The authors have done extensive study on the previous work, out of that some of the most relevant work to the current research are discussed in this section.

Padwalkar et al. [3] presents a hybrid technique of secure multi-party computation, in this authors uses random number for data privacy, in the hybrid protocol participating parties and third party contribute for computation so it will be faster. This paper does not deal with communication security threat, when some parties are targeted purposefully and the case when third party becomes malicious.

Ayday et al. [5] presents a privacy preserving system for storing and processing genomic, clinical and environmental data by using privacy preserving integer comparison and homomorphic encryption. In this DNA sequence of patient is created by certified institution using the sample provided by the patient. The clinical and environmental data of the patient is collected from doctor, patient visits, or could be directly provided by the patient. (For ex. age weight, family history by patient whereas cholesterol level blood sugar level by his/her doctor's visits). All these information is considered as sensitive and need to be protected. Here, SMC is applied to preserve privacy of patients, against curious parties at storage and processing unit (SPU) and malicious parties at medical unit (MU). Genome is next big thing in medical science to identify disease risks. It could be possible by ensuring privacy of patients' sensitive data during the tests.

Sheikh et al. [7] explains the importance of modified ck-secure sum protocol over ck-secure sum protocol. In this protocol initiator changes its position in unidirectional ring, so that no neighbour remains together for more than one round. Here, data is divided into n segments (where 'n' is the number of party). On $n^{th}$ round initiator announces the sum. For this protocol to work effectively minimum four parties are required. It doesn't deal with malicious parties and security threat.

Sheikh et al. [8] presents dk-secure sum protocol using ring arrangement. In this paper, parties exchange any 1 out of k segments, with any 1 party out of k, so all the parties have k-1 segment, plus 1 received from other party. In this protocol if two parties collaborate to get third parties data, then it may break the protocol (in case of three parties). This protocol works efficiently for four or more parties. In this paper authors assumed that communication channels are secure, it doesn't deal with insecure communication channels.

Sheikh et al. [9] presents a ck-secure sum protocol, in the proposed work authors divided data into fixed segments. The authors claim that, here probability of data leakage is zero. As this secure sum protocol uses changing neighbour mechanism, where neighbours are changed in each round with fixed protocol initiator. For this protocol to work accurately, minimum four parties are required. It doesn't deal with malicious behaviour of party and attacker during communications.

Clifton et al. [11] gives tools for privacy preserving data mining; in this random number mechanism is used to preserve privacy of individuals. In this, if two parties collaborate they can get the data of third party.

## III. PROBLEM DEFINITION

The secure sum of parties' personal inputs is a good example of SMC which has attracted the attention of researchers from organization and academics to develop SMC protocol with minimum data leakage and higher confidentiality and security.

The secure sum protocol was initiated by Clifton et al. [11] in this authors used randomization technique for joint computation. In the proposed protocol participating parties were organized in a one-way ring. One party works as originator of the protocol through which computation begins by deciding a random number and adding it to its private data. The sum is forwarded to next party for further computation and so on.

Privacy preservation for secure multi-party computation problems have been achieved by other methods also [10, 6, 12] by mean of randomization and cryptography. But these solutions suffer from privacy loss in certain scenario. This paper presents simulation of protocol for privacy preservation to solve secure sum problem using randomization and anonymization.

### A. Formalization of Secure Sum Problem

The secure sum problem consists of set of parties who wish to perform collaborative sum over parties' private input without revealing the data and identity of participants.
The SS problem is as follows:
1. A set of n parties $P_1, P_2 \ldots P_n$ each party holds private data $x_i$ where i $\in$ (1, 2… n).
2. A personal data set D = $\{x_1, x_2 \ldots x_n\}$ is the set of all the parties' personal data.
3. A secure sum of parties personal data need to be computed without disclosing sensitive information.
4. The inter party constraints exists between the parties to share the data among the parties (if required).
5. The intra-party constraints of party $P_i$ on $x_i$ are the personal information of party $P_i$. It is known only to party $P_i$.
6. A reasonable solution $S_{ln}$ is a representation of the secure sum of variable set.

## IV. SIMULATION OF SECURE SUM PROTOCOL

This section presents a set of simulation analysis of the secure sum protocol. The privacy and security levels of the protocol has been analyzed using probabilistic analysis, here main emphasis is on the performance. Here the questions are: 1) how does this protocol work in case of colluding anonymizers? 2) How does this secure sum protocol perform in various settings, and how does this achieve the basic objective of privacy of sensitive information of individual participants?

This paper demonstrates the implementation of the randomization and anonymization based protocol for secure sum using time bound random function to generate the inputs. In order to answer first question, experiment has been performed ensuring no anonymizer get more than one packet from the same party so in this case probability of getting more than one packet of same party by malicious anonymizers is insignificant. To answer second question, the data is synthetically generated with varying parameters to test the protocol in different settings. Table 1 presents the parameters used for simulation. As shown in Fig. 1, for execution in batch these parameters are fetched from the text file which contents the values for these parameters.

In this paper experiment is performed in following settings:

Protocol initiator will decide number of party, packets per party and anonymizers are decided such that $m \geq (n \times (t_{pk} + t_{pk}))/m_x$ here '$n$' is the number of party, '$t_{pk}$' is the packets per party, and '$m_x$' is the maximum limit of an anonymizer. All the experiments have been executed considering there is only one TTP.

### A. Simulation Setup

The experiments have been executed using IntelCore2 duo T5450 processor clocked at 1.66 GHz and 2 GB of memory under windows 7 operating system. The secure sum protocol is evaluated using different set of parameters as shown in Table 1. For the test run parties input is generated using time based randomization function. All the result presented, are averaged from 500 test runs in batches. The tests runs in batches of 500 runs each, marking the average of the times noticed in the runs.

TABLE I. Table of Simulation Parameters

| Parameter Name | Description | Default Value |
|---|---|---|
| $n$ | Number of participating parties ( minimum number of parties are 2) | 10 |
| $m$ | Number of Anonymizers (minimum number of anonymizers are 4) | $m \geq (n \times (t_{pk} + t_{pk}))/m_x$ |
| $t_{pk}$ | Number of Packets per party ( minimum 3) | 3 |

### B. Distributed Randomized Secure Sum (DRSS)

In the protocol [4] only one random number is added to the data so probability of breaking the protocol would be (1/n+1) but in this protocol it is reduced as different random numbers are added to each packet. The protocol can break only when all the random numbers and data packets are joined correctly. So the probability of breaking the protocol would be (1/n)*(1/n) which is insignificance and very less compare to

(IJCSIS) International Journal of Computer Science and Information Security,
Vol. 12, No. 11, 2014JCRA [4]. This protocol was proposed by Shukla et al. [1] in this protocol at first the personal data is divided into packets, parties use dummy data to hide personal data contained in each packet. Hence, each party in this protocol must be able to divide the data into packets and generate dummy data using time bound random function. Than all the encoded packets are send to arbitrarily selected anonymizers. Anonymizers forward the packets to data pool and random number pool respectively. Finally it is the responsibility of TTP to compute the final secure sum using data and random number pool.

### *A.A.1 Simulation Steps*

In the simulation *DRSS* protocol runs in the batches, the input data and random numbers are synthetically generated using random function, with varying parameters to test the protocol in different settings. After completion of the each run, outcome is stored for analysis & testing purpose. Once a batch run is complete the average of the time for each run is taken for comparative analysis.

## V. RESULT AND ANALYSIS

### *A. Data Outflow probability*

"Fig. 1" shows the experimental analysis, it illustrates that the data is divided by the party in the form of packets and distributed to different anonymizers. It is shown in the graph, if randomization factors are increased data leakage probability is reduced to negligible, it means the individual data privacy is increased.

$$\Pr(l,m) = \left(\frac{l}{m}\right)^{(2k)} \quad (1)$$

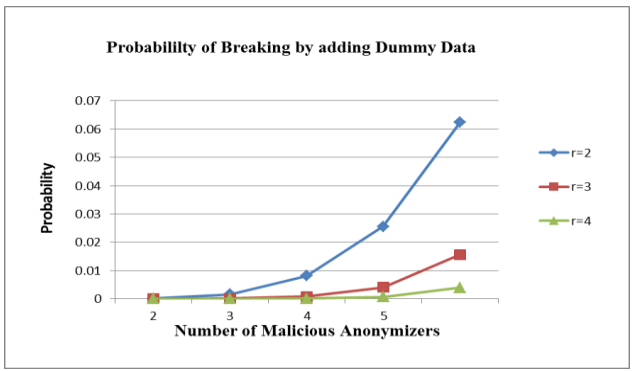

Figure 1 Data outflow probability by increasing randomization factor

### *B. Time Complexity*

To find the impact of the proposed algorithm, a simulator was designed in Visual C++ and corresponding time complexities were recorded. The simulation result presents the proposed techniques although increases the time complexity but guarantees security in communication and computations. Graphs shown below illustrate the experimental results of execution time by increasing number of parties, packets and anonymizers respectively.

The time complexity is theoretically evaluated considering best, average and worst cases. The theoretical results are shown in "Fig. 2".

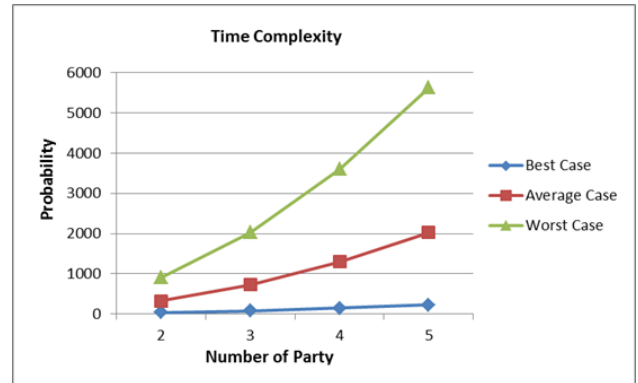

Figure 2 Result of theoretical time complexities

*a) Case 1: Increasing Number of Party*

"Fig. 3" shows the result, when the number of party increases, the overall computation time increases. It includes the time, for randomization of each party's data and packet distribution to different anonymizers. The test result shows that protocol is following the rule.

TABLE II. Test Result by increasing number of party

| Number of Party | Average time taken in batch of 500, for tpk=3, m=5 |
|---|---|
| 2 | 761.3353293 |
| 3 | 770.3306122 |
| 4 | 777.428 |
| 5 | 785.788 |
| 6 | 795.916 |
| 7 | 797.916 |
| 8 | 806.67 |

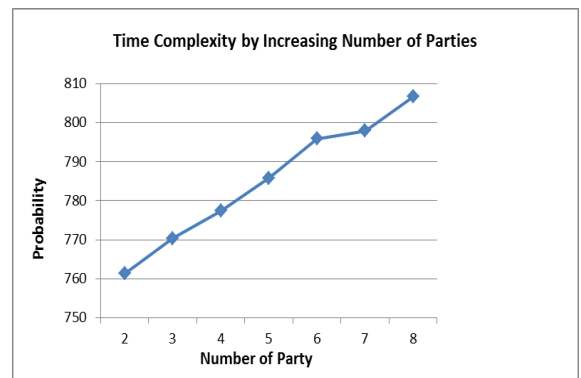

Figure 3. Time Complexity by increasing parties

*b) Case 2: Increasing Number of Packets and Party together*

"Fig. 4" shows that increase in number of packets, increases time complexity, but as the number of packets increases privacy and security improves. As per the constraint no anonymizer get same parties packet more than once, to achieve this, it is required that the number of anonymizers should always be equal to or greater than the total number of packets per party. (Here m=5)

TABLE III. Test Result by increasing number of packets as well as parties

| Number of Parties ➡ Packets ⬇ | 3 | 4 | 5 |
|---|---|---|---|
| 2 | 787.877 | 816.298 | 865.072 |
| 3 | 798.3306122 | 837.692 | 867.356 |
| 4 | 837.064 | 840.082 | 873.272 |
| 5 | 867.974 | 927.8808081 | 1017.392 |

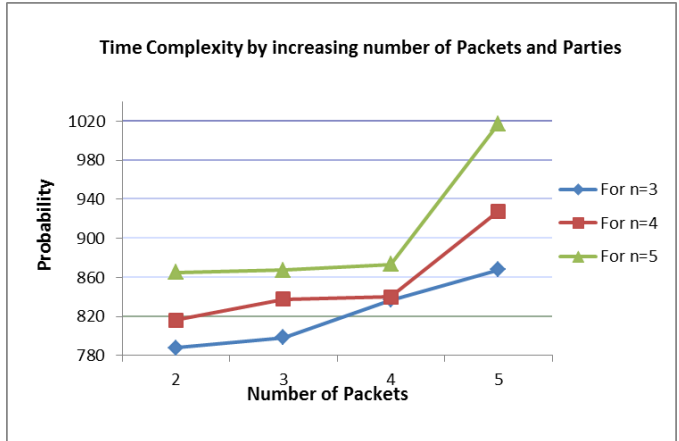

Figure 4. Time Complexity by increasing packets and parties together

*c) Case 3: Increasing Number of Anonymizers*

"Fig 5." shows that as the number of anonymizers increases, the execution time reduces so time complexity reduces. It reduces the overall cost of computations.

TABLE IV. Test Result by increasing number of anonymizers

| Number of Anonymizers | For n=3, $t_{pk}$=3 |
|---|---|
| 5 | 798.3306122 |
| 6 | 773.32 |
| 7 | 763.802 |
| 8 | 758.782 |

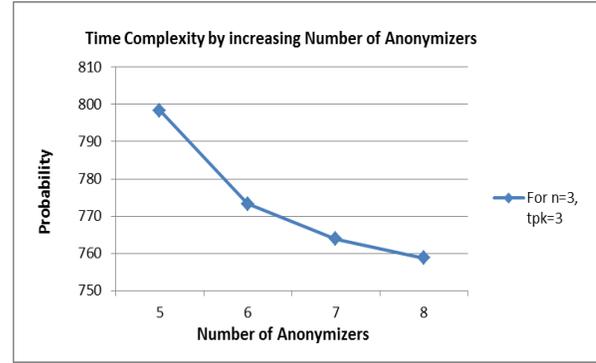

Figure. 5 Time Complexity by increasing Anonymizers

*C. Computation Complexity*

In the proposed protocol overall computation complexity at TTP is constant as in final computation TTP need to perform only one operation to generate the secure sum of multiple parties.

## VI. CONCLUSIONS AND FUTURE WORKS

This paper shows the simulation result of a secure sum protocol. This result shows that, this is a noticeable improvement to already existing protocols. It provides better privacy, security against the hackers attack during communications. As compare to other protocols currently simulated protocol perform better, in case of semi-honest party for minimum of three parties. These results in high privacy, security and confidentiality which are highly important in medical, banking, and industries. In future the data distribution algorithm can be optimized to enhance the performance.

ACKNOWLEDGMENT

Authors wish to acknowledge Mr. Kapil Tiwari, EMC[2], Bangalore, India for his technical guidance and support.

REFERENCES

[1] Samiksha Shukla, Dr G Sadashivappa, "A Distributed Randomization Framework for Privacy Preservation in Big Data", presented in CSIBIG "978-1-4799-3064-7/14/$31.00 ©2014 IEEE", 8th – 9th March'2014.
[2] Samiksha Shukla, Dr G Sadashivappa, "Secure Multi-Party Computation Protocol using Asymmetric Encryption" in Proceedings of the 8th INDIACom; INDIACom-2014, International Conference on "Computing for Sustainable Global Development", organized by Bharati Vidyapeeth's Institute of Computer Applications and Management (BVICAM), New Delhi (INDIA), 5th – 7th March, 2014.
[3] Anand R Padwalkar, Prajakta Pande and Vidhi Dave, Secure Multi-Party Computation Protocol: Basic Buidling Block Methods, National Conference on Innovations in IT and Management : 2014, SIMCA, Pune, India, ISBN: 978-81-927230-0-6, pp-128, 21st -22nd February'2014.
[4] Samiksha Shukla and Sadashivappa G, An Algorithm for SMC with analysis of malicious conduct, International Journal of Advanced Research in Computer Science and Software Engineering, ISSN: 2277 128X, Volume 3, Issue 10, 667-673, October 2013.
[5] Ayday, Erman, et al. "Privacy-Preserving Computation of Disease Risk by Using Genomic, Clinical, and Environmental Data.", 2013.
[6] Kamara, Seny, Payman Mohassel, and Mariana Raykova. "Outsourcing Multi-Party Computation." IACR Cryptology ePrint Archive, pp-272, 2011.

[7]   Sheikh, Rashid, Beerendra Kumar, and Durgesh Kumar Mishra. "A Modified ck-Secure Sum Protocol for Multi-Party Computation." arXiv preprint arXiv:1002.4000, 2010.

[8]   Sheikh, Rashid, Beerendra Kumar, and Durgesh Kumar Mishra. "A distributed k-secure sum protocol for secure multi-party computations." arXiv preprint arXiv:1003.4071, 2010.

[9]   Sheikh, Rashid, Beerendra Kumar, and Durgesh Kumar Mishra. "Changing Neighbors k Secure Sum Protocol for Secure Multi Party Computation." arXiv preprint arXiv:1002.2409, 2010.

[10]  D. K. Mishra, M. Chandwani. "Extended Protocol for Secure Multiparty Computation using Ambiguous Identity. "WSEAS Transaction on Computer Research, vol. 2, issue 2, February 2007.

[11]  C. Clifton, M. Kantarcioglu, J. Vaidya, X. Lin, and M. Y. Zhu, "Tools for Privacy-Preserving Distributed Data Mining," *J. SIGKDD Explorations, Newsletter,* vol.4, no.2, ACM Press, pages 28-34, December 2002.

[12]  Lindell, Yehuda, and Benny Pinkas. "Privacy preserving data mining." *Journal of cryptology* 15, no. 3, 177-206, 2002.

## AUTHORS PROFILE

### MS SAMIKSHA SHUKLA

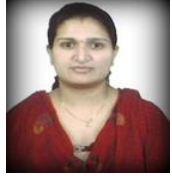

Ms Samiksha Shukla received the M.Tech degree in Computer Science and Engineering from RGTU, Bhopal, India. Presently she is pursuing PhD in CSE from Christ University, Bangalore, India. She has around a decade of teaching and research experience. She has published around 30 papers in refereed International/National Journals and Conferences including IEEE. She is the program committee member of several conferences. She is an associate member of Computer Society of India.

### DR. G. SADASHIVAPPA

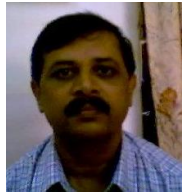

G. Sadashivappa received the BE degree in Electronics Engineering from Bangalore University in 1984 and M.Tech degree in Industrial Electronics from KREC ( NIT-K), Mangalore University in1991. During 1984-1989 he worked as Lecturer in AIT Chickmagalore, JMIT Chitradurga and Engineer trainee in Kirloskar Electric Co Ltd, Unit-IV, Mysore. Since 1992 is working in R.V.College of Engineering, Bangalore. He obtained his Doctoral degree from VTU Belgaum during 2011 in the area of image processing. His research areas include Image & Video Coding, Biomedical Signal Processing, underwater communication and network protocols with data security.

### DR. DURGESH KUMAR MISHRA

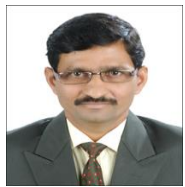

Dr. Durgesh Kumar Mishra has received M.Tech degree in Computer Science from DAVV, Indore in 1994 and PhD in Computer Engineering in 2008. Presently he is working as Professor (CSE) and Director Microsoft Innovation Centre at Sri Aurobindo Institute of Technology, Indore, MP, India. He is having around 24 Yrs. of teaching experience and more than 6 Yrs. of research experience. His research topics are Secure Multi-Party Computation, Image processing and cryptography. He has published more than 80 papers in refereed International/National Journals and Conferences including IEEE and ACM. He is a senior member of IEEE, Computer Society of India and ACM. He has played very important role in professional society as Chairman. He has been a consultant to industries and Government organization like Sales tax and Labor Department of Government of Madhya Pradesh, India.